# Thermal and electrical conductivity of a refractory high-entropy alloy after high-pressure torsion: Electron versus phonon contributions

Jacqueline Hidalgo-Jiménez[1], Payam Edalati[2,3], Md Amirul Islam[1], Makoto Arita[4], Bidyut Baran Saha[1], Kaveh Edalati[1,*]

[1] WPI, International Institute for Carbon Neutral Energy Research (WPI-I2CNER), Kyushu University, Fukuoka 819-0395, Japan
[2] School of Mechanical Engineering, FEM, University of Campinas, UNICAMP, Campinas, Brazil
[3] School of Applied Sciences, University of Campinas, FCA-UNICAMP, Limeira Brazil
[4] Department of Materials, Faculty of Engineering, Kyushu University, Fukuoka 819-0395, Japan

The equiatomic refractory high-entropy alloy TiZrHfNbTa was processed by high-pressure torsion (HPT) to investigate the effect of nanostructuring and defect engineering on thermal and electrical transport properties. Severe plastic deformation (SPD) via the HPT treatment induces substantial accumulation of dislocations, grain refinement to the nanometer level (average: 40 nm), and partial transformation from the BCC phase to the ω phase. While hardness increases to a steady state with processing, the specific heat capacity exhibits a non-monotonic behavior: it decreases at low strains due to the suppression of low-frequency vibrational modes by dislocations, then partially recovers at high strains due to anharmonic vibrations at newly formed high-angle grain boundaries. Thermal conductivity decreases at low strains but shows a slight recovery at high strains, whereas electrical conductivity decreases monotonically to a steady state without recovery. Analysis using the Wiedemann-Franz law reveals that the electronic contribution dominates thermal transport, while the phononic contribution (limited by the scattering of phonons on defects) is only 11 to 23%, depending on the degree of straining. The contrasting evolution of thermal and electrical conductivity is ascribed to the transition from dislocation-dominated vibrations at low strains to grain boundary-dominated vibrations at high strains, which affects phonons and electrons with different efficiencies.

**Keywords:** high-entropy alloys (HEAs); lattice defects; phononic scattering; electronic scattering; heat transport

*Corresponding author (email: kaveh.edalati@kyudai.jp)

## 1. Introduction

High-entropy alloys (HEAs), also referred to as compositionally complex materials or multi-principal element alloys, have emerged as an important class of metallic materials because they depart from the traditional alloy-design approach based on one dominant element [1,2]. Instead, HEAs contain several principal elements in substantial or near-equiatomic concentrations, often leading to high configurational entropy over 1.5*R* (*R*: gas constant) in the solid state [1]. This compositional concept can stabilize simple solid-solution phases while simultaneously introducing severe lattice distortion, complex chemical interactions through the cocktail effect and unusual defect and diffusion behavior [3]. As a result, HEAs have attracted extensive attention for their outstanding combinations of mechanical strength, fracture resistance, strength-ductility synergy, wear performance and other structural properties [4]. Beyond structural applications, growing interest now focuses on the use of HEAs and their partner high-entropy ceramics in energy systems, cryogenic technologies, catalysis, hydrogen storage, superconductivity, biomedical devices and other functional applications, where their chemically tunable nature offers major advantages over conventional alloys [5].

Among the many HEAs, refractory HEAs are particularly attractive because they are based on high-melting-point elements and can maintain useful performance under harsh thermal and mechanical conditions [6]. Many refractory HEAs crystallize predominantly in a body-centered cubic (BCC) structure, which is favored by the atomic size and electronic characteristics of early transition elements like titanium, zirconium, hafnium, niobium, tantalum, molybdenum, tungsten, etc. [7]. The BCC structure is technologically important because it can provide high specific strength, thermal stability and compositional flexibility, while also enabling interesting functional responses linked with electronic structure and phase stability [8]. The equiatomic TiZrHfNbTa alloy has become one of the most representative BCC refractory HEAs [9] and has been studied as a model system for understanding deformation mechanisms [10], corrosion behavior [11], hydrogen interactions [12], superconductivity [13] and biomedical performance [14]. It was shown that properties of the alloy is significantly affected by nanostructural design [15].

Severe plastic deformation (SPD) techniques provide a powerful route for such nanostructural design because they can impose exceptionally large strains while preserving overall sample integrity [16]. SPD processing leads to vacancy generation [17], dislocation accumulation [18], nonequilibrium grain boundary formation [19], deformation-induced phase transformations [20], defective interphase stabilization [21] and, more importantly, extreme grain refinement in various materials [22]. These features have been widely exploited to strengthen metals, improve fracture toughness, modify corrosion behavior, enhance hydrogen storage, and improve functional responses in various alloys and compounds [22]. Amongst the available SPD technologies, high-pressure torsion (HPT) is one of the most effective because it combines torsional shear with high hydrostatic pressure [23], enabling processing not only conventional metals but also hard-to-deform materials such as biomolecules [24], ceramics [25] and refractory HEAs [26]. While HPT has been applied to biomolecules to understand the origin of life [24], a main focus of HPT processing of ceramics has been the synthesis of functional compounds like high-entropy photocatalysts for hydrogen production [27], $CO_2$ conversion [28] and antibiotic degradation [29]. In refractory HEAs, recent studies demonstrated that HPT can enhance superconducting performance in TiZrHfNbTa through nanostructuring and defect generation [30], and can simultaneously improve strength and boost biocompatibility [31]. Nevertheless, despite these advances, no systematic study has clarified how SPD processing affects the electrical conductivity, thermal conductivity and related thermophysical properties such as specific heat capacity of refractory HEAs, including TiZrHfNbTa. This is a critical gap because transport properties govern the potential use of refractory materials.

The current study, therefore, aims to investigate the influence of HPT-induced nanostructuring on the electrical and thermal transport phenomena in the TiZrHfNbTa high-entropy alloy. By correlating structural evolution with hardness, specific heat capacity, thermal conductivity and electrical conductivity, the study clarifies the competing roles of lattice defects. The results reveal that heat capacity and both thermal and electrical conductivity are reduced by HPT-induced defects at early stages of deformation, but their thermal properties are slightly recovered by a transition from dislocation accumulations to high-angle grain boundaries. The findings also establish that electron scattering is the dominant phenomenon for the transport behavior of the refractory HEA.

## 2. Experimental Procedures

Equiatomic TiZrHfNbTa was synthesized through arc melting of high-purity elemental titanium, zirconium, hafnium, niobium and tantalum (purity >99.9%) under a purified argon atmosphere. To minimize chemical segregation and ensure compositional homogeneity, the ingot was flipped and remelted seven times. The solidified alloy was subsequently sectioned into discs using electrical discharge machining. Typical disc dimensions were ~1 cm in diameter and 0.8 mm in thickness. Prior to SPD processing, the discs were homogenized at an elevated temperature of 1373 K for 60 min and then quenched in water (to be referred to as $N = 0$ sample in figures). The homogenization target was to obtain a chemically uniform coarse-grained BCC solid solution and to eliminate potential effects of electrical discharge machining on the microstructure.

The process of SPD was conducted by HPT (order-made by Riken Enterprise, Japan) under a quasi-constrained condition. Each disc was placed between upper and lower anvils containing shallow circular depressions and processed at ambient temperature under a pressure of 6 GPa. Torsional straining was introduced at a rate of 1 rpm for selected turns of $N = 1$ and 10. The shear strain ($\gamma$) was estimated from the standard HPTstrain relationship using radial distance ($r$), turns ($N$) and disc thickness ($h$): $\gamma = 2\pi rN / h$ [32]. Examination of the slippage using a method described earlier [33] showed that no slippage between the sample and the surfaces of HPT anvils occurred, suggesting that shear strain is effectively applied during the process.

Phase analysis was performed through X-ray diffraction (XRD) using Cu Kα radiation (SmartLab, Rigaku, Japan). Diffraction patterns were recorded over an appropriate $2\theta$ diffractometer angle range with a small step size of 0.05º to resolve peak broadening and possible minor phases. Peak positions were used to evaluate phase constitution, whereas peak broadening was used to assess defect generation by the HPT treatment.

Vickers microhardness test was carried out on polished disc surfaces using a microhardness tester (HM-103, Mitutoyo, Japan). Indentations were placed at four radial locations from the center of the samples in order to capture the strain gradient inherent to HPT processing. A regular force of 0.5 kgf and an indentation time of 15 s were used for all samples. Measurements were conducted at different radial distances as a sign of microstructural evolution.

Microstructural characterization was performed by transmission electron microscopy (TEM) under an acceleration voltage of 200 kV (JEM-ARM200F, JEOL, Japan). Thin foils were prepared from regions near 3.5 mm from the sample center, where the imposed strain represents the average imposed strain along radii. Discs with a 3 mm diameter were cut using electric discharge machining, ground to a reduced thickness of 150 µm, followed by twin-jet electropolishing and final ion milling until electron transparency was achieved. Different detectors for bright-field imaging, dark-field imaging, selected-area electron diffraction (SAED), high-resolution imaging and fast Fourier transform (FFT) analyses were used to identify nanograins, dislocations, grain boundaries and deformation-induced phase transformations.

Specific heat capacity ($C_P$) was determined using a heat flux-type differential scanning calorimeter (DSC-60A, Shimadzu Corporation, Japan) specifically designed for heat capacity measurements, featuring a baseline noise level below 0.5 μW and a certified limit of permissible relative error of ±5%. Small specimens with 30–50 mg mass were cut from the processed discs and heated alongside a reference sample at a heating rate of 5 $Kmin^{-1}$. The heat flow difference between the sample and reference was monitored over the selected temperature range. Heat capacity values were derived from calibration procedures and the slope of the heat-flow response. To ensure reproducibility, at least five repeated measurements were performed for each condition, and the relative standard deviation was calculated to be smaller than ±5%.

Thermal diffusivity ($\alpha$) was determined by laser flash analysis (LFA 457, NETZSCH, Germany). The surface of the polished discs was covered with graphite using a spray, and after drying, the disc was inserted into the machine. A short laser pulse was applied to one face of the specimen and the transient temperature rise on the opposite face was detected by an infrared sensor. Measurements were conducted over a controlled temperature range from ambient temperature to 330 K. Thermal conductivity ($\kappa$) was determined using thermal diffusivity ($\alpha$), density ($\rho$ = 9850-9950 kg $m^{-3}$, measured by the Archimedes method) and specific heat capacity: $\kappa = \alpha\rho C_P$ [34].

Electrical conductivity measurements were carried out at ambient temperature using a conventional four-point probe technique (homemade device). Rectangular bars with 0.6×0.6 $mm^2$ cross-sectional area ($A$) and 9 mm length were cut from discs, while the radial positions of the centers of all bars were at 2 mm to maintain a consistent strain history. The sample surfaces were polished prior to testing to ensure good electrical contact, which was maintained during the measurements using spring-loaded probes. A constant current of $I$ = 0.2 A was applied through the outer probes while the voltage between the two inner probes, located at $l$ = 4 mm away from each other, was recorded. The electrical conductivity ($\sigma$) of the central region of bars, which corresponded to shear strains of $\gamma$ = 0 for the homogenized sample, $\gamma$ = 16-22 for the $N$ = 1 sample and $\gamma$ = 157-222 for the $N$ = 10 sample, was calculated as the reciprocal of electrical resistivity through the standard equation: $\sigma = 1 / ((V/I)(A/l))$ [35].

**3. Results**

Fig. 1a exhibits the XRD patterns of TiZrHfNbTa in the homogenized condition and following the HPT treatment for 1 and 10 turns. The homogenized sample exhibits sharp diffraction peaks indexed to a single-phase BCC with the (110), (200), (211), (220) and (310) reflections, consistent with previous reports on this alloy [9-15]. No additional peaks corresponding to secondary phases are detected in the homogenized state, confirming the successful synthesis of a BCC solid solution. Following the HPT treatment for 1 and 10 turns, the BCC peaks remain present but become noticeably broader. The full width at half maximum (FWHM) of the (110) BCC peak is plotted versus HPT turns in Fig. 1b. The FWHM increases progressively from approximately 0.55° in the homogenized sample to 0.75° after 10 turns. This broadening reflects the continuous refinement of the microstructure and the accumulation of lattice defects, particularly dislocations, in agreement with reported results for other HPT-treated materials [16-22].

Fig. 1c shows the hardness versus radial distance from the disc center for samples treated with different HPT turns. The homogenized sample exhibits a uniform hardness of approximately 330 Hv across the entire disc diameter. Following the HPT treatment for $N = 1$ turn, a clear hardness gradient develops, with values increasing from 330 Hv at the central region to ~460 Hv at the peripheral region ($r = 4$ mm). This gradient reflects the strain distribution inherent to HPT, where shear strain increases linearly with radial distance [23,32]. After $N = 10$ turns, the hardness becomes more uniform across the disc and reaches a saturation value of approximately 460 Hv. The microhardness data are replotted versus shear strain in Fig. 1d, where the microhardness rises sharply with enhancing strain in the initial deformation stage, followed by a more gradual increase until saturating at a steady state. This saturation behavior is characteristic of SPD-processed materials, where a steady-state microstructure is attained through a dynamic balance between dislocation generation and recovery processes as well as a balance between grain boundary generation, migration and recrystallization [36].

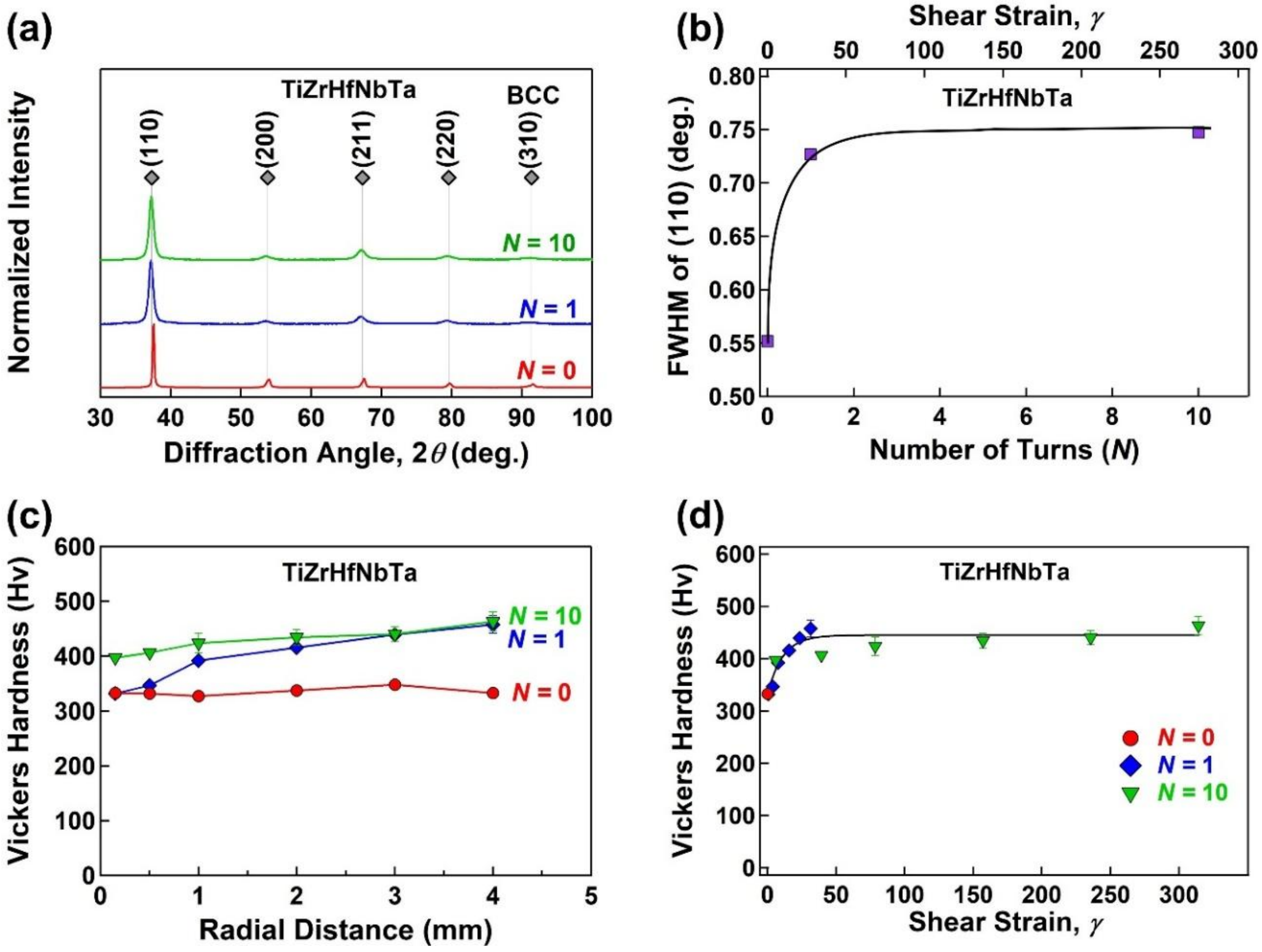

Fig. 1. (a) XRD profiles, (b) FWHM for (110) peak versus HPT turns, (c) hardness versus radial distance and (d) hardness versus shear strain for TiZrHfNbTa high-entropy alloy after homogenization ($N = 0$) and HPT treatment for 1 and 10 turns.

Fig. 2 shows TEM bright-field images, dark-field images and SAED diffractograms of the TiZrHfNbTa alloy after (a) homogenization and following the HPT treatment for (b) 1 and (c) 10 turns. The initial material exhibits a coarse-grained structure after homogenization. Bright-field and dark-field TEM images of the homogenized sample show a single grain with a [110] zone-axis SAED pattern, where the palm tree-like features correspond to dislocation substructures formed during water quenching after homogenization. After $N = 1$ turn, the microstructure consists of submicrometer grains, although grain boundaries are not well-defined. The SAED diffractogram exhibits a ring shape, demonstrating the presence of ultrafine grains. After $N = 10$ turns, a fully nanocrystalline microstructure is developed, and grain boundary borders become more distinct, with a mean grain size of approximately 40 nm, as illustrated in Fig. 2d. The SAED diffractograms after 10 turns shows continuous and uniform rings, confirming the random crystallographic orientation of the nanograins and the existence of a high percentage of high-angle grain boundaries, a feature which is frequently reported in HPT-treated materials including HEAs [26,30,31].

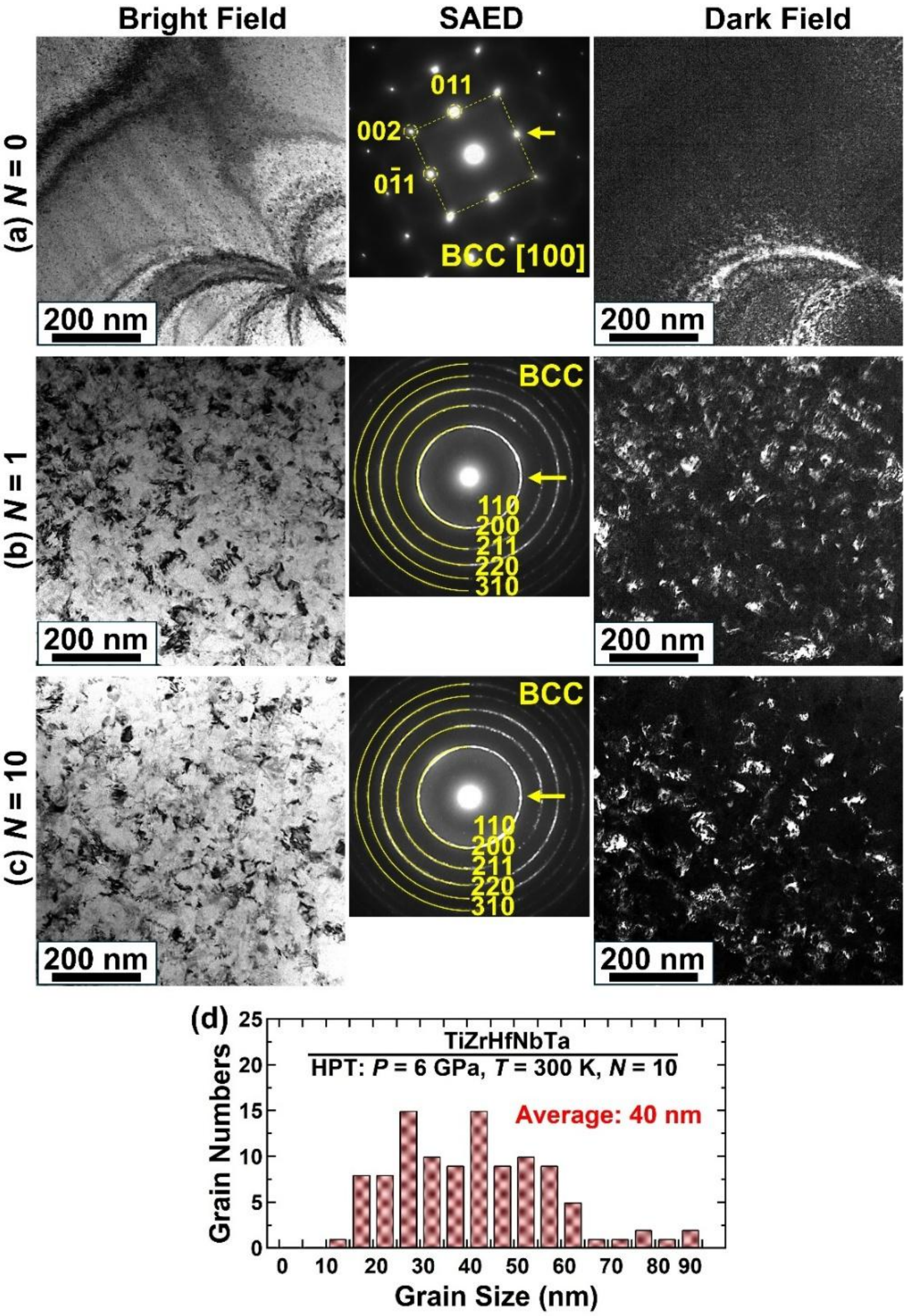

Fig. 2. TEM bright-field micrographs, SAED diffractograms and dark-field micrographs for TiZrHfNbTa high-entropy alloy after (a) homogenization, (b) HPT processing for 1 turn and (c) HPT processing for 10 turns. (d) grain size histogram for sample treated for 10 turns.

Fig. 3 presents high-resolution TEM images and corresponding FFT diffractograms of the sample treated using HPT for (a-e) 1 and (f, g) 10 turns. Fig. 3a reveals the presence of what appears to be a BCC grain. However, close examination of the FFT profiles of this region, as shown in Fig. 3b-d, indicates the existence of an ω structure nanocrystal embedded within the BCC matrix. The ω phase is a primitive hexagonal structure that forms through a displacive transformation in titanium, zirconium, hafnium and their alloys. While the ω phase is an equilibrium phase in these metals under high pressure, it can remain metastable under ambient pressure after HPT processing. The ω phase has been observed in HPT-treated titanium [37], zirconium [38], hexagonal close-packed (HCP) titanium alloys [39] and BCC titanium alloys [40]. One difficulty with the evaluation of this phase is that it can hardly be detected by XRD due to the overlap of its peaks with the BCC phase. Besides the presence of a new phase, dislocations are frequently observed, as illustrated in Fig. 3e. After $N$ = 10 turns, high-angle grain boundaries are repeatedly observed within high-resolution images, indicating a transition from dislocation activity to grain boundary stability. Transition from dislocation activity to high-angle grain boundary formation is a general feature of severely deformed materials [16,22].

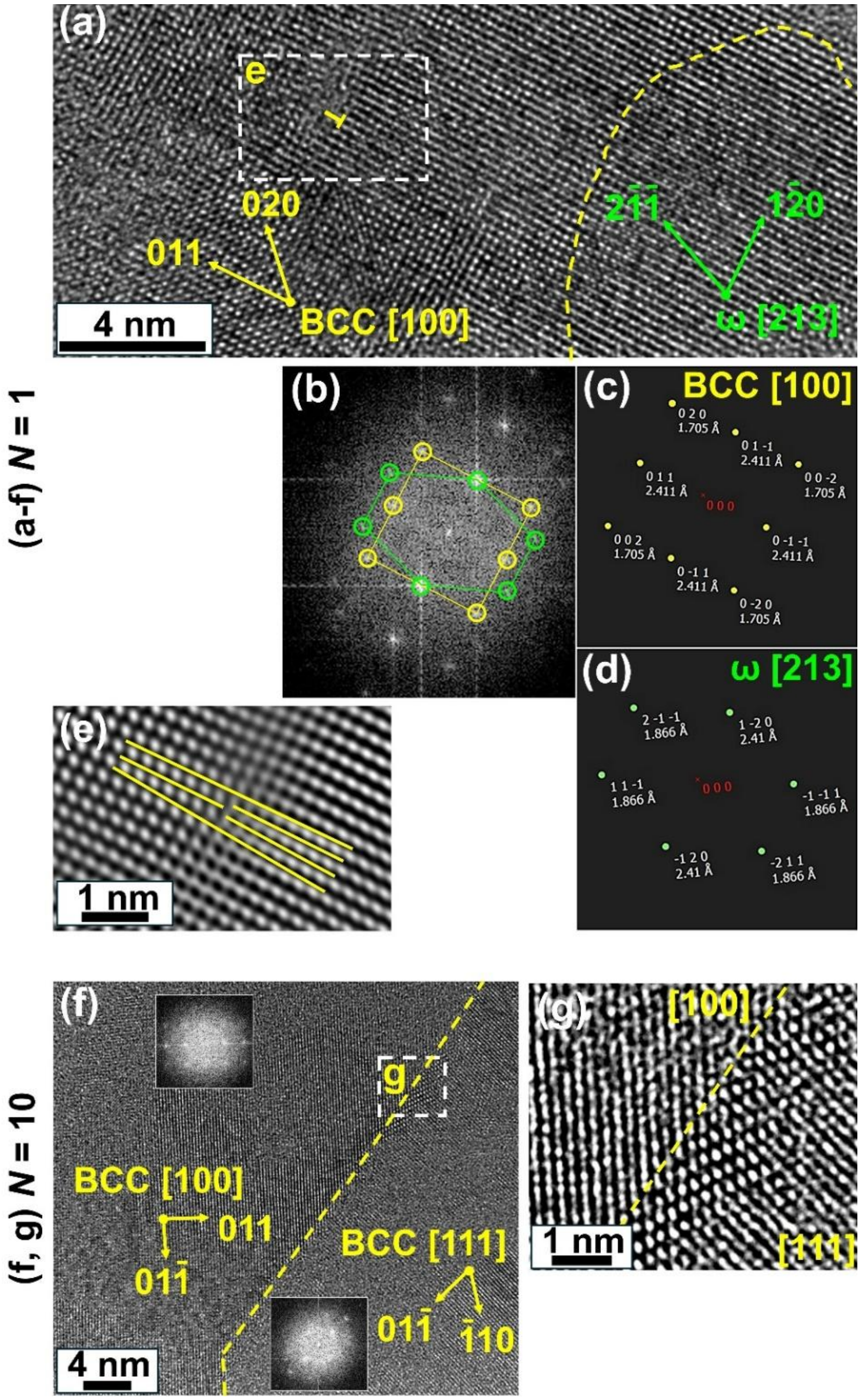

Fig. 3. High-resolution TEM images for TiZrHfNbTa high-entropy alloy following HPT treatment for (a-f) 1 turn and (f, g) 10 turns, where (a) is high-resolution images, (c) is FFT profile taken from right side of image in (a), (d) is simulated FFT pattern of ω phase from [213] zone axis, (e) is lattice image of dislocation from area indicated in (a), (f) is high-resolution image, and (g) is lattice image of grain boundary from area indicated in (f).

Fig. 4a exhibits the specific heat capacity versus temperature for the homogenized sample and samples treated using HPT for 1 and 10 turns. Specific heat capacity values in this plot are 249-254 $JK^{-1}kg^{-1}$ for the homogenized sample, 210-211 $JK^{-1}kg^{-1}$ for the $N = 1$ sample and 228-230 $JK^{-1}kg^{-1}$ for the $N = 10$ sample. For all samples, the specific heat capacity increases with increasing temperature, as expected from the progressive activation of phonon vibrational modes [35]. At a given temperature, the homogenized sample exhibits the highest heat capacity, while the HPT-processed samples show lower values. The evolution of specific heat capacity at 318 K against HPT turns is shown in Fig. 4b. The specific heat capacity decreases from the homogenized state to 1 turn, indicating that the introduction of lattice defects initially reduces the ability of the material to store thermal energy per unit mass. However, after further processing to 10 turns, the specific heat capacity shows an increase compared to 1 turn, although it remains below the homogenized value. This non-monotonic behavior suggests competing effects: at low to intermediate strains, defect and dislocation accumulation reduces heat capacity, while at very high strains, the formation of a stable nanograined microstructure with high-angle grain boundaries partially recovers the heat capacity. Although there are apparently no earlier studies on the heat capacity of HPT-treated materials, non-monotonic behaviors against strain have been occasionally reported for some properties, such as the hardness of aluminum [41] or copper [42].

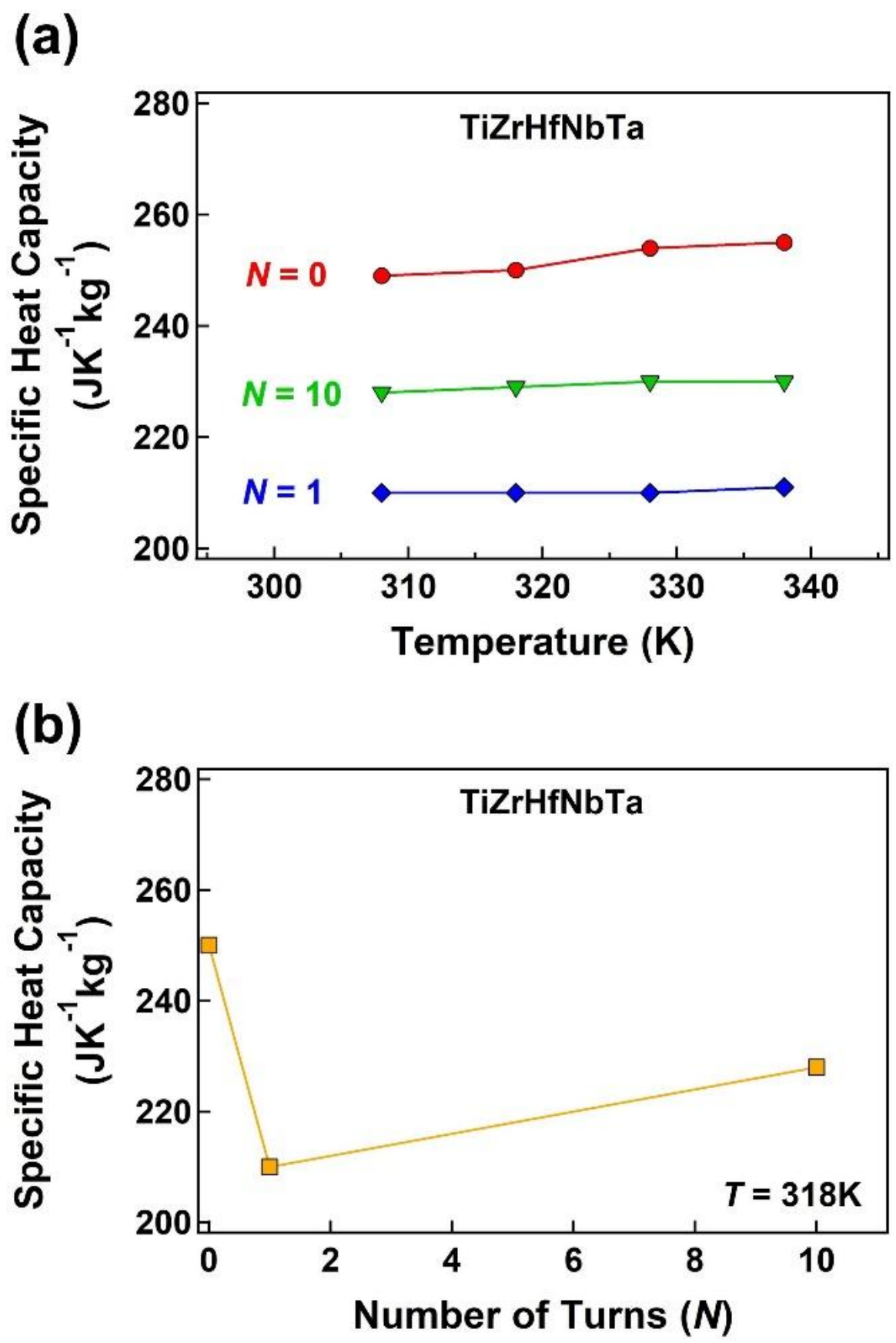

Fig. 4. Specific heat capacity versus (a) temperature, (b) number of HPT turns for TiZrHfNbTa high-entropy alloy after homogenization and HPT treatment for 1 and 10 turns.

Fig. 5a presents the thermal conductivity versus temperature for the homogenized sample and samples treated using HPT for 1 and 10 turns. Thermal diffusivity values in this plot are 4.46-4.78 $mm^2s^{-1}$ for the homogenized sample, 4.31-4.53 $mm^2s^{-1}$ for the $N = 1$ sample and 4.10-4.39 $mm^2s^{-1}$ for the $N = 10$ sample. For all samples, thermal conductivity increases with temperature, a behavior typical of disordered alloys where phonon scattering is dominant [34]. The variation of thermal conductivity at 318 K with HPT turns is illustrated in Fig. 5b. The homogenized sample has the highest thermal conductivity, with a value of approximately 10.9 W m$^{-1}$·K$^{-1}$. Following the HPT treatment for 1 turn, thermal conductivity decreases, reaching approximately 8.9 W m$^{-1}$·K$^{-1}$. This decline can be simply ascribed to increasing electron and phonon scattering by deformation-induced defects [43]. After 10 turns, the thermal conductivity increases slightly compared to $N$ = 1, reaching approximately 9.3 W m$^{-1}$·K$^{-1}$, despite a decrease in thermal diffusivity and due to an increase in specific heat capacity. This recovery, though modest, suggests that the transition from a dislocation-dominated microstructure to a nanograined structure with high-angle grain boundaries affects vibrational modes, possibly due to changes in the nature of defect-phonon and defect-electron interactions [44].

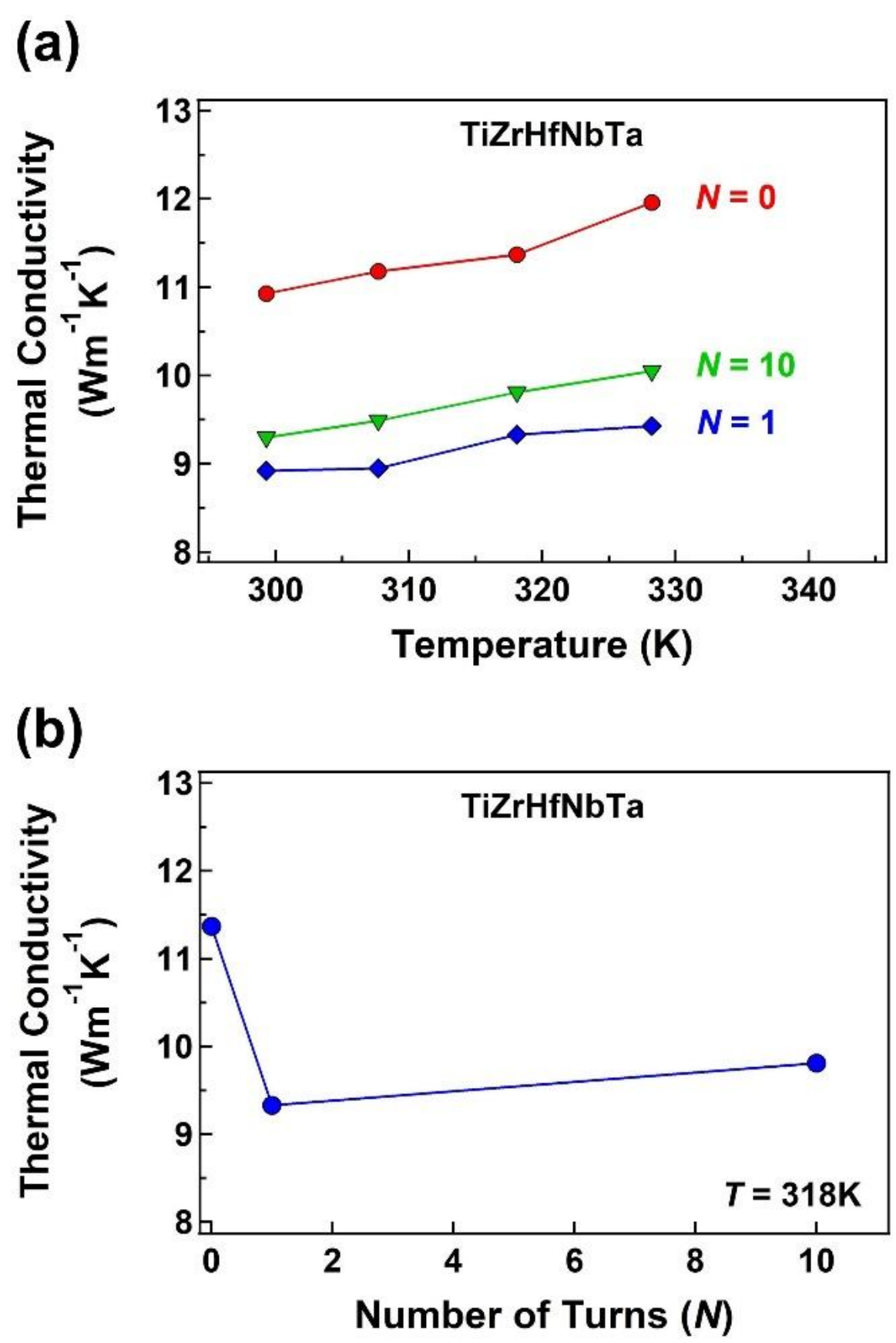

Fig. 5. Thermal conductivity versus (a) temperature, (b) number of HPT turns for TiZrHfNbTa high-entropy alloy after homogenization and HPT treatment for 1 and 10 turns.

Fig. 6 shows the electrical conductivity at ambient temperature versus HPT turns. Electrical resistivity values in this plot are 0.87 μΩm for the homogenized sample, 0.92 μΩm for the $N$ = 1 sample and 0.93 μΩm for the $N$ = 10 sample. The homogenized sample exhibits the highest electrical conductivity, with a value of approximately 1.15 MS m$^{-1}$. Upon HPT processing for 1 turn, the electrical conductivity decreases substantially to approximately 1.08 MS m$^{-1}$, reflecting increased electron scattering by lattice defects. With further processing to $N$ = 10 turns, the electrical conductivity shows a very small decrease and remains almost at a steady-state level. The alteration of electrical conductivity with the HPT turns is consistent with that reported for other severely deformed materials such as copper [45] and aluminum alloys [46]. Unlike thermal conductivity, electrical conductivity did not show a slight recovery at $N$ = 10, indicating that electrons and phonons respond differently to the evolving defect structure.

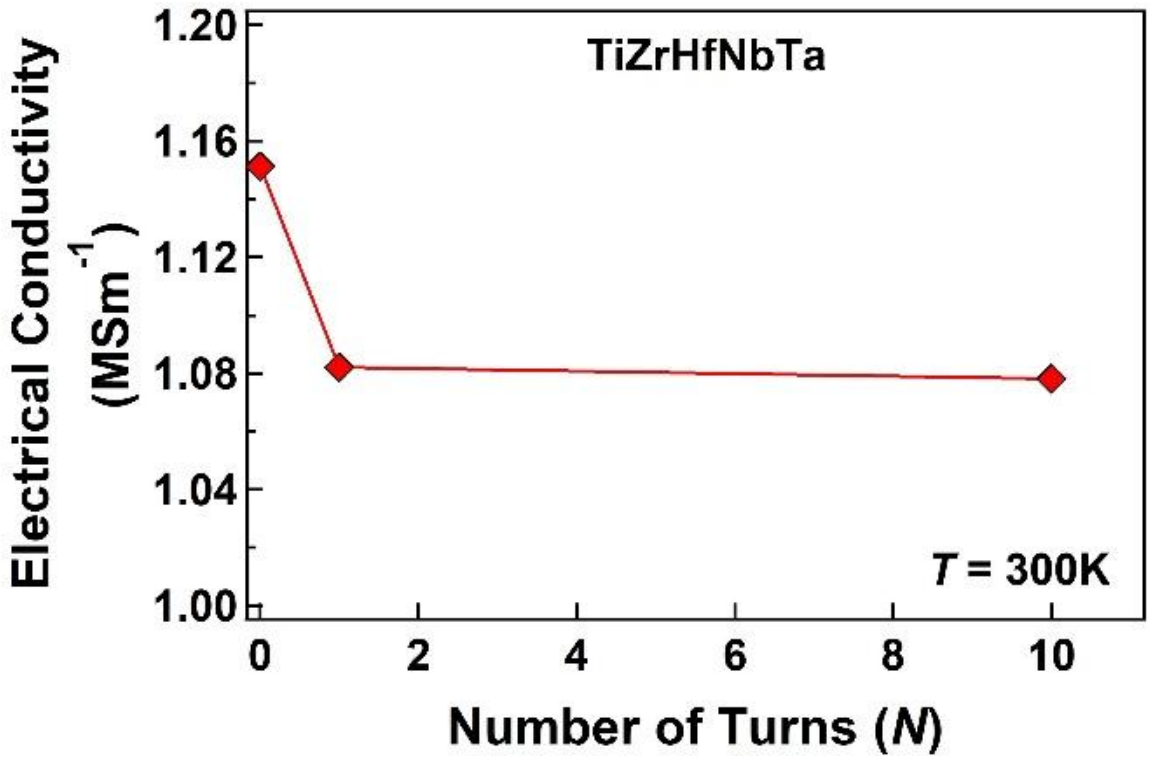

Fig. 6. Electrical conductivity versus number of HPT turns for TiZrHfNbTa high-entropy alloy after homogenization and HPT treatment for 1 and 10 turns.

**4. Discussion**

Understanding SPD effects on thermal and electrical transport in HEAs is essential not only for their functional applications but also to understand electron-defect and phonon-defect interactions in this emerging group of materials [1-5]. By considering the importance of transport phenomena in refractory materials and increasing interest in their functional applications [6-15], a refractory HEA was selected for this study. In the BCC TiZrHfNbTa alloy, the transport response to HPT processing follows an expected behavior at the early stages of deformation, with reductions in both thermal and electrical conductivity. However, a partial recovery of thermal conductivity at high strain is observed, which is not expected. These observations reflect the complex interplay between defect generation, grain refinement, vibrational modes, and electron and phonon scattering, as two major contributors to transport phenomena [35].

The observed trends can be explained by considering how different types of lattice defects interact with electrons and phonons. Dislocations, which are abundant after $N$ = 1 turn, create long-range strain fields that effectively scatter both electrons and phonons, leading to the sharp reductions in thermal and electrical conductivity at this stage [47,48]. At $N$ = 10 turns, the microstructure transforms to nanograins with high-angle grain boundaries. Grain boundaries scatter electrons strongly due to the disruption of the periodic potential, which explains why electrical conductivity does not recover [43-46]. However, grain boundaries may affect the vibrational modes and specific heat capacity and scatter phonons less efficiently than dislocations because phonons with wavelengths longer than the grain boundary thickness can transmit across interfaces [49,50]. This difference in scattering efficiency allows the phonon contribution to thermal conductivity to increase slightly at $N$ = 10, producing the modest recovery observed in Fig. 5b. This recovery is not due to enhancement of thermal diffusivity, but it is due to an increase in the specific heat capacity. The changes in transport phenomena in the presence of dislocations and grain boundaries have been reported in some earlier studies as well [44,51].

Using the Wiedemann-Franz law, the electronic contribution ($\kappa_e$) to conductivity can be assessed from electrical conductivity at a given temperature of $T$ through $\kappa_e = L_0 T\sigma$ , where $L_0$ = $2.44 \times 10^{-8}$ W·Ω·K$^{-2}$ is the Lorenz number [35]. The phonon contribution ($\kappa_p$) is then assessed by subtracting the electronic contribution from the total thermal conductivity through $\kappa = \kappa_e + \kappa_p$ [35] Fig. 7 presents the (a) absolute and (b) relative contributions of electron scattering and phonon scattering to the total transport for the homogenized sample and after HPT processing. For all conditions, transport of heat by electrons dominates, contributing approximately 77 of the total thermal conductivity of the homogenized sample, while transport of heat by phonons contributes the remainder. This dominance of electronic thermal transport is characteristic of metallic alloys with high electrical conductivity and a high density of free electrons [35]. With HPT processing for $N = 1$ turn, the absolute values for both electronic and phonon contributions decrease due to enhanced scattering by lattice defects. The electronic contribution is still the dominant mechanism with approximately 89% contribution. A further increase in the HPT turns to $N = 10$ does not make an appreciable change in the electronic contribution, but the phonon contribution slightly increases. This indicates that the unusual slight recovery of thermal conductivity after $N = 10$ is due to the phonon vibrational modes and scattering by transition from dislocations to grain boundaries, while electron scattering remains high after such microstructural transition [44,51].

To further assess the transition in phonon vibrational modes, the phonon mean free path ($\Lambda_p$) can be determined from kinetic theory [35]: $\kappa_p = C_P \rho v_s \Lambda_p / 3$ ($C_P$: specific heat capacity per mass unit under constant pressure, $\rho$: density, and $v_s$: sound velocity, which is 3028 m s$^{-1}$ for this HEA [52]). As shown in Fig. 7c, the phonon mean free path remains well below the Casimir limit of ~40 nm dictated by the grain size, suggesting the intrinsic strong phonon scattering by chemical disorder in this high-entropy alloy. The phonon mean free path reduces from 1.0 nm in the homogenized state to 0.5 nm after 1 HPT turn, confirming the strong additional scattering by deformation-induced dislocations. After 10 HPT turns, the phonon mean free path partially recovered to 0.6 nm, demonstrating that a transition from dislocations to high-angle grain boundaries affects the vibrational modes of phonons. This observation is consistent with prior studies showing that phonon scattering by grain boundaries can be less efficient than by dislocation strain fields [44,51]. It should be noted that, although the volume fraction of the ω phase could not be quantified, its formation and the associated BCC/ω interphase boundaries are expected to introduce additional scattering sites that would reduce, rather than enhance, thermal conductivity. This supports the attribution of the partial recovery at $N = 10$ turns primarily to the dislocation-to-grain boundary transition rather than to the ω phase.

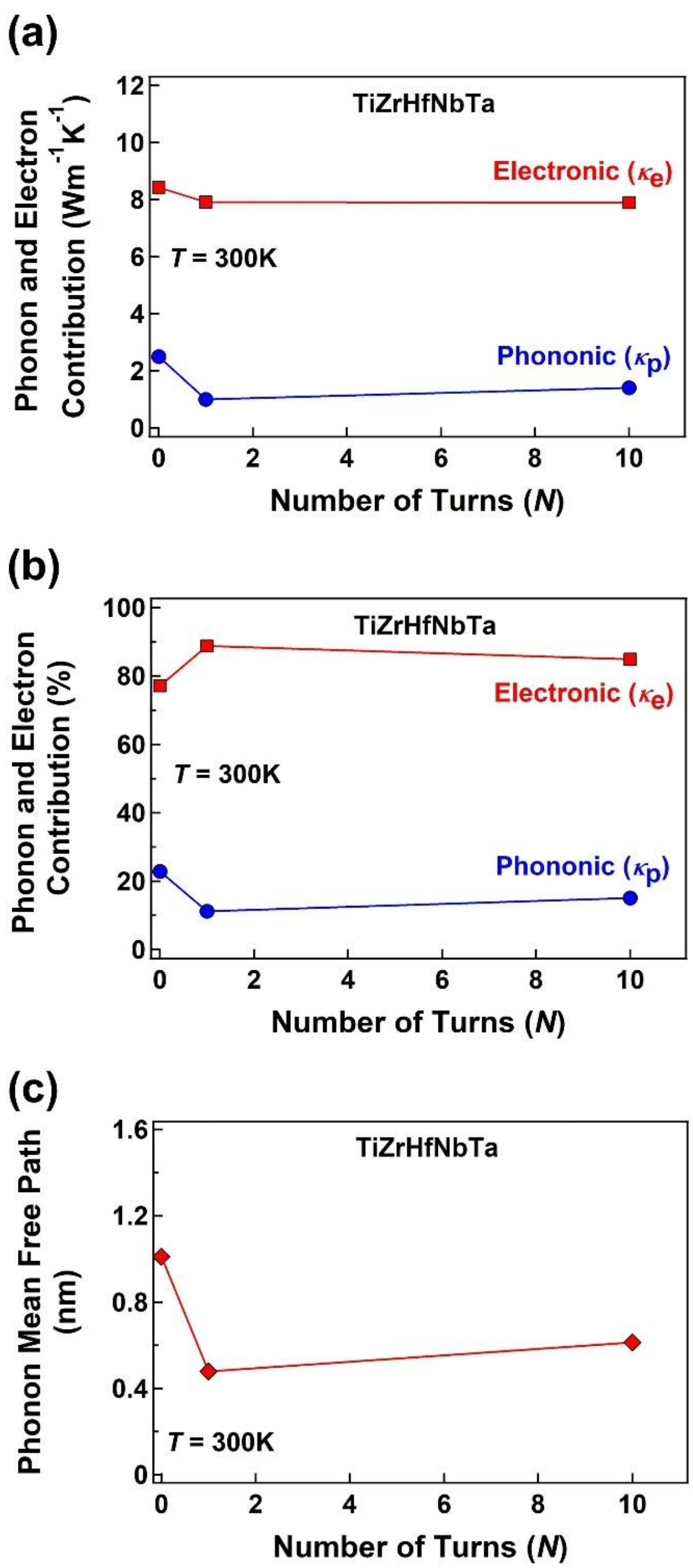

Fig. 7. Contribution of electrons and phonons to thermal conductivity in (a) absolute values and (b) percentage, and (c) phonon mean free path for TiZrHfNbTa high-entropy alloy after homogenization and HPT treatment for 1 and 10 turns.

Finally, it should be noted that the decrease in specific heat capacity at $N = 1$ reflects the suppression of low-frequency vibrational modes by deformation-induced defects, particularly dislocations, which locally reduce the density of phonon states available for thermal activation [47,53]. The partial recovery at $N = 10$ can be due to the generation of a stable nanograined microstructure with high-angle grain boundaries, where excess free volume and anharmonic vibrations at the interfaces introduce additional phonon modes, thereby increasing the heat capacity [54,55]. This non-monotonic behavior indicates that different defect types have opposing effects on vibrational modes and transport phenomena in HEAs. Future studies should clarify the exact contributions of HPT-induced defects and nanostructuring on transport phenomena by detailed theoretical studies and by extending this experimental study to a wider range of HEAs with different crystal structures, including refractory HEAs [6-14,56,57]. Moreover, simultaneous examination of the temperature dependence of electrical and thermal conductivity, as well as systematic study of the effect of the number of HPT turns (including only compression) on microstructure and corresponding conductivity, can further clarify the transport phenomena in HEAs. This study can also be extended to other SPD methods, reported in the literature [58,59], to clarify the effect of deformation mode on thermal conductivity of HEAs.

## 5. Conclusions

This study demonstrates that severe plastic deformation can decouple thermal and electrical transport in a refractory high-entropy alloy by inducing different scattering efficiencies for dislocations and grain boundaries. While dislocations effectively attenuate both electrons and phonons, high-angle grain boundaries maintain strong electron scattering but allow changes in phonon vibrational modes, leading to a partial recovery of thermal conductivity without improving the electrical conductivity. The non-monotonic evolution of specific heat capacity further reveals that dislocations suppress low-frequency vibrational modes, whereas grain boundaries introduce anharmonic contributions that enhance vibrational entropy. These findings establish that nanostructuring through high-pressure torsion provides a viable route to tailor thermal and electrical transport properties in refractory high-entropy alloys, offering new opportunities for applications requiring combinations of strength and specific heat and electrical performance.

## CRediT Authorship Contribution Statement

Jacqueline Hidalgo-Jiménez: Conceptualization, Methodology, Investigation, Validation, Writing – review & editing; Payam Edalati: Conceptualization, Methodology, Investigation, Validation, Writing – review & editing; Md Amirul Islam: Conceptualization, Methodology, Investigation, Validation, Writing – review & editing; Makoto Arita: Conceptualization, Methodology, Investigation, Validation, Writing – review & editing; Bidyut Baran Saha: Conceptualization, Methodology, Investigation, Validation, Writing – review & editing; Kaveh Edalati: Conceptualization, Methodology, Investigation, Validation, Writing – review & editing.

## Declaration of competing interest

The authors declare no competing interests that could have affected the presentation in this manuscript.

## Acknowledgements

The authors thank Prof. Dmitry Chernyshov of the European Synchrotron Radiation Facility, Grenoble, France, for the discussion. This research was supported partly by the Japan Science and Technology Agency (JST) through the ASPIRE project (JPMJAP2332).

## Data Availability

The data will be available for sharing upon contact with the corresponding author.